# Title:
# Nano-to-micro spatiotemporal imaging of magnetic skyrmion's life cycle

**Authors:**
T. Shimojima[1,*] A. Nakamura[1], X. Z. Yu[1], K. Karube[1], Y. Taguchi[1], Y. Tokura[1,2,3], K. Ishizaka[1,2]

**Affiliations:**
[1]RIKEN Center for Emergent Matter Science (CEMS), Wako 351-0198, Japan
[2]Quantum-Phase Electronics Center (QPEC) and Department of Applied Physics, The University of Tokyo, Tokyo 113-8656, Japan
[3] Tokyo College, The University of Tokyo, Tokyo 113-8656, Japan

*Corresponding author

**Magnetic skyrmions are the self-organized topological spin textures behaving like particles. Because of their fast creation and typically long lifetime, experimental verification of skyrmion's creation/annihilation processes has been challenging. Here we successfully track skyrmions dynamics in defect-introduced $Co_9Zn_9Mn_2$, by using pump-probe Lorentz transmission electron microscope. Following the nanosecond-photothermal excitation, we resolve 160-nm-skyrmion's proliferation at <1 ns, contraction at 5 ns, drift from 10 ns to 4 µs and coalescence at ~5 µs. These motions relay the multiscale arrangement and relaxation of skyrmion clusters in a repeatable cycle of 20 kHz. Such repeatable dynamics of skyrmions, arising from the weakened but still persistent topological protection around defects, enables us to visualize the whole life of the skyrmions, as well as demonstrating the possible high-frequency manipulations of topological charges brought by skyrmions.**

**Introduction**

Skyrmion is a topologically stable particle first proposed in nuclear physics(*1*). It is characterized by a topological charge representing an integer how many times the field configuration wraps a sphere. Recently it has been proven that this concept can be applied to condensed matter physics(*2-5*). Indeed, previous Lorentz transmission electron microscopy (LTEM)(*6*) studies revealed winding spin textures in non-centrosymmetric chiral magnets, referred to as magnetic skyrmions. In these materials, the competition between the relativistic Dzyaloshinskii–Moriya (DM)(*7,8*) and ferromagnetic exchange interactions stabilizes the triangular skyrmion lattice (SkL) under the external magnetic field. Here, the energy scale of atomic exchange energy *J* is required for creation (annihilation) of a skyrmion. When *J* is much larger than the thermal fluctuation of $k_BT$, it in general ensures the skyrmion's long lifetime at a certain temperature *T*, that is associated with the topological protection(*9*).

The diameter of skyrmions in chiral magnets typically ranges from ~5 to ~100 nm depending on the competition of the magnetic interactions. Since the size is much larger than the lattice constant, the skyrmions are free from the commensurate pinning, and thus obtain high mobility(*9*). When there are any imperfections in the system, which inevitably exist as defects, disorders, edges, interface roughness, *etc* in real materials, skyrmions tend to deform(*10,11*) and become rather flexible due to the weakened topological protections(*12*). Understanding the effect of such imperfections on skyrmion dynamics has been an important issue in skyrmion-based applications (*13,14*). The combination of the flexible particle character with high internal degrees of freedom is expected to bring about the multiscale dynamics analogous to the soft matters(*15,16*). Although skyrmion dynamics such as creation/annihilation(*17,18*), deformation and self-vibrations(*19,20*), drift motion(*21,22*), and many-body interactions(*23,24*) have been independently investigated by specific experimental probes, how these dynamics sequentially evolve in the multiple spatiotemporal scales is not yet clarified. This issue should be highly relevant in operating the skyrmion-based spintronic devices(*9,13,14*), where the kinetics of skyrmions are playing the pivotal roles for the performance. Nevertheless, its comprehensive investigation has been a challenging issue, particularly because of difficulties in achieving a wide spatiotemporal range down to nanoscale for the magnetic imaging techniques.

Here, we report nano-to-micro spatiotemporal imaging of the skyrmions in a chiral magnet $Co_9Zn_9Mn_2$ thin film with defects introduced. We use pump-probe LTEM (Fig. 1A) with high spatiotemporal resolutions of 10 nm and 10 ns. By irradiating the nanosecond-photothermal pulses, we continuously resolve proliferation at <1 ns, contraction at 5 ns, drift from 10 ns to 4 μs and coalescence at ~5 μs which relay the multiscale ordering and relaxation of skyrmions cluster in a repeatable cycle of 20 kHz. The flexibility and short lifetime of the skyrmions are attributed to the weakened but still persistent topological protection around the local defects. Our observations visualize the unique life cycle of the photothermally-induced metastable skyrmions around the defects. While majority of the past research on magnetic alloys had been rather focusing on clean skyrmion crystals, our present result offers important aspects in skyrmion physics, which are associated with the major processes of manipulating skyrmions.

**Results**
**Pump-probe LTEM on metastable skyrmions**

For the pump-probe LTEM measurements, we prepared the partially disordered skyrmions condensed state by introducing crystal defects. First, by employing the focused ion-beam method, we shaped 100-nm thin plate from single crystalline $Co_9Zn_9Mn_2$(*25,26*). Thin-plate $Co_9Zn_9Mn_2$ exhibits a helical spiral magnetic structure and triangular SkL in the equilibrium phase diagram (Fig. 1B), as represented by the magnetization textures obtained by analyzing LTEM images with a transport-of-intensity equation at 370 K under 0 mT and 35 mT magnetic fields, respectively (Fig. 1, C and D). The magnetic modulation period (~160 nm) is determined by $2\pi J/D$, where $D$ is the DM interaction. With field cooling, the equilibrium skyrmion phase persists down to room temperature (RT) as a metastable state (Fig. 1E) where the skyrmions show long lifetime(*27*) whose temperature dependence can be described by the Arrhenius law(*28,29*). Here, we irradiated the thin plate with Ga ions to randomly introduce crystal defects [see Materials and Methods of the supporting online material (SOM)]. Such non-magnetic defects are known to induce deformation of the magnetic structures(*10-12*). In this Ga-ion irradiated sample, as shown in Fig. 1F, we first observed the locally distorted helical spirals at RT. We then irradiated the 1 ns pulsed laser (20 kHz for 5 seconds) with the fluence of $F$ = 5.1 mJcm$^{-2}$ which induces rapid temperature cycles between RT and ~380 K (SOM, section 1). The field cooling delivers the locally distorted metastable skyrmions at RT (Fig. 1G), some of them appearing as the elliptical skyrmions around defects. In the present pump-probe LTEM measurements, we rather focus on these skyrmions which are expected to exhibit flexible transformations with short lifetime upon excitation due to the weakened but still persistent topological protection(*12*) (SOM, section 2). In this experimental setup, we can extract the essentially "repeatable" dynamics of skyrmions within 50 µs by integrating 36 million pump-probe cycles in one image (20 kHz for 30 minutes acquisition time). As will be discussed later, this methodology is highly effective for investigating the intrinsic mean motions of the skyrmions, while smearing out the rare events(*30,31*).

**Overviewing one-cycle dynamics of metastable skyrmions**

In the pump-probe measurements, we homogeneously irradiated the thin plate with nanosecond pulsed laser with the fluence of $F = 5.1$ mJcm$^{-2}$ to induce the skyrmion dynamics. First, we estimate the change in the sample (lattice) temperature(*32*) caused by the laser irradiation, from the transient intensity of the Bragg spots (Fig. 2A) obtained by the time-resolved electron diffraction measurements (Fig. 2B). The rapid decrease at time zero and the exponential recovery with a time constant of 1.4 µs represent the temperature jump up to 380 K in 1 ns (pulse duration of pump laser) and the subsequent relaxation back to RT in 1.4 µs, respectively (thick black curve in Fig. 2B).

The LTEM images before and after the temperature jump show that the disordered (locally 7 bonded) skyrmion state (–430 ns, Fig. 2C) changes into the quasi-hexagonal skyrmion cluster (270 ns, Fig. 2D), suggesting the photo-thermal induced disorder-order transformation. A series of the LTEM images in Fig. 2, E1 to E8, (taken from the region depicted by broken-line squares in Fig. 2, C and D) indicate that the magnetic contrast gradually transforms as a function of delay time *t*, showing the one-cycle dynamics of the skyrmions. Transient sample temperatures are indicated at the bottom of each panel. Here we focus on adjacent two elliptical skyrmions (~350 nm × 160 nm) presented in Fig. 2E1 ($t = -15$ ns), colored by blue (skA) and red (skB) in Fig. 2F1, at RT under 17 mT magnetic field. As shown in the series of LTEM images (Fig. 2E) and the schematics (Fig. 2F), skA and skB exhibit different but correlated dynamics. Within the time resolution, skB splits into two smaller and more isotropic skyrmions (skB1 and skB2 with diameters of ~160 nm), *i.e.* skyrmion proliferation. On the other hand, skA rapidly shrinks in 10 ns accompanying the shift of the center of mass, and then realizes the nearly circular shape of ~160 nm. After that, skB1 moves upward for ~100 ns, to where the distances among the skyrmions become uniform thereby forming a rather isotropic skyrmion cluster. These subsequent dynamics in Fig. 2, E1 to E5, can be interpreted as the self-organization process associated with the disorder-order transformation. After ~300 ns, on the other hand, the relaxation process sets in, and the skB1 starts to move back to its initial position with slight deformation. Finally, skB1 and skB2 recombine to form the elliptical skyrmion around ~5000 ns (skyrmion coalescence) and skA simultaneously elongates to recover its initial shape. Almost identical magnetic patterns at –15 ns (Fig. 2E1) and 7820 ns (Fig. 2E8) proves that the observed skyrmion dynamics are repeatable in this time scale. Hereafter, we closely look at individual skA and skB with quantitative analysis of the transient LTEM images.

**Skyrmion proliferation, reshaping and self-organization toward ordered cluster (0 - 300 ns)**

First, we show the initial change of the skyrmion caused by the thermal jump from RT to 380 K. Figure 3, A1 and A2 exhibit the LTEM images obtained at –10 ns and 10 ns, respectively which indicate a suppression of the magnetic contrast after photoexcitation. Detailed *t* dependence of its spatial integration as shown in Fig. 3D suggests that the suppression reflects the ultrafast demagnetization process. It has been known that the demagnetization caused by the jump in the electron temperature occurs in a short time [e.g. ~100 fs for ferromagnetic alloy(*33*)]. Although the present observation is naturally limited by the time resolution (gray area in Fig. 3D), we can define the time zero as the reference to the subsequent slower skyrmion dynamics as follows.

Next, we discuss the skyrmion proliferation and reshaping processes occurring in 20 ns from time zero. To investigate the proliferation of skB, we show the intensity profiles of the LTEM image along line 1 (Fig. 2E1). As shown in Fig. 3B, an additional signal shows up in the middle of skB around time zero (double headed arrows B → B1 and B2). It corresponds to the emergence of the magnetic wall in skB indicated by the black arrows in Fig. 2, E3 and F3, suggesting the proliferation process of skB (~330 nm) to skB1 (~160 nm) and skB2 (~160 nm). Detailed $t$ dependence of the magnetic wall formation, estimated by the intensity in the white dashed rectangle in Fig. 3B, suggests that the proliferation occurs concurrently with the demagnetization around time zero (see gray area in Fig. 3E). On the other hand, skA contracts to form a circular shape. In the line-profile of the LTEM image along line 2 (Fig. 2E1) in Fig. 3C, the contraction seems to be completed in ~10 ns (see double headed arrow A). We quantitatively analyze the width of skA ($W_A$) from the contour plots in Fig. 2G, obtained from the LTEM images in Fig. 2E (SOM, section 3). Detailed $t$ dependence of $W_A$ indicates that the shrinkage from ~310 nm to ~160 nm occurs ~5 ns behind the demagnetization and skyrmion proliferation (gray area in Fig. 3F). These results reveal the proliferation (time-resolution limited) and contraction (5-ns delayed) processes of the flexible skyrmions, which realize the circular skyrmions with a diameter comparable to that at equilibrium ~380 K (160 nm in Fig. 1D).

Difference in the response time of these two processes can be understood by considering underlying spin dynamics. The skyrmion proliferation can be regarded as the topological transformation accompanied by the skyrmion nucleation. In general, once the local magnetic moment is reversed by external stimuli with an energy scale of $J$, the DM interaction sequentially aligns neighboring spins, and finally form the skyrmion as a winding spin texture(*9,34*). Numerical simulation suggested that such skyrmion nucleation typically completes in sub-ns by a pulse-laser heating(*35*), which is consistent with the resolution-limited proliferation around time zero (Fig. 3E). Regarding the retardation behavior in the contraction, the theoretical studies focusing on the skyrmion motion associated with the internal spin degrees of freedom(*36,37*) may help understanding. Here, when a skyrmion is driven by a time-dependent external force, the internal spins start precessions giving rise to a large effective mass and delayed drift motion(*37*). Numerical simulation(*35*) demonstrated that the spin orientations in the skyrmion are highly disturbed just after the pulse-laser heating. The disorder in the spin orientation is then reduced showing a damped oscillation to form stable and smooth winding texture. The damping time (retardation time) can be several orders of magnitude larger than the time required for skyrmion nucleation (sub-ns), depending on the Gilbert damping constant of the material(*35*). We can thus expect that the delayed response in the skyrmion contraction (shift of the center of mass) is caused by the peculiar response in the internal degrees of freedom which is detectable in the present time resolution.

After acquiring the isotropic circular shapes, the skyrmions rather seem to move as particles. Analysis on the contour plots for skB/skB1 indicates the shift along $b$ ($S_B$) as functions of time. We found that skB1 starts to move behind the contraction of skA (gray areas in Fig. 3, F and G), which could be naturally understood as the skB1 moving toward a space produced around skA. The increase in $S_B$ continues until ~100 ns. We found that at 100 - 300 ns where $S_B$ takes maximum (orange areas in Fig. 3, H and I), the quasi-hexagonal skyrmion cluster is realized around skA (Fig. 2D), in contrast to the initial state (–430 ns) where skA is surrounded by seven distorted skyrmions (Fig. 2C). Observed skyrmion cluster rearrangement is consistent with an attractive force between skyrmions which can drive skyrmion-cluster formation in a helical background(*24*).

These results indicate that the proliferation, contraction, and drift motion of respective skyrmions (in ~10 ns, ~100 nm scales) are well correlated, and develop into the better-ordered skyrmion cluster of larger spatiotemporal scales (~300 ns, sub-μm).

**Relaxation process and skyrmion coalescence (300 ns - 10 μs)**

After ~300 ns, the quasi-hexagonal skyrmion cluster starts to relax back to the initial disordered state. As shown in Fig. 4, A and B, the relaxations of $W_A$ and $S_B$ in the $t < 4$ μs region are reproduced by assuming the exponential decay with a time constant obtained from the transient temperature in Fig. 4C (1.4 μs) (SOM, section 4). Around 4 - 6 μs, on the other hand, the system shows a complicated behavior and sudden recovery to the initial state. To investigate the "mean" dynamics of the skyrmions around 5 μs, we analyze the intensity profile along line 3 (Fig. 2E3). Figure 4D exhibits the splitting of the intensity across the time zero, indicated by the double headed arrow B1 representing the skB1 part, suggesting the proliferation process of skB to skB1 and skB2 as abovementioned. The size of proliferated skB1 remains constant until ~4 μs (Fig. 4D), and collapses at around 5 μs (skyrmion coalescence) as shown in Fig. 4E. Such discontinuity should be reflecting the glimpse of the "topological protection" by which the skyrmion annihilation cannot be achieved by the adiabatic deformation. Considering that the sample temperature is settled back close to ~300 K already at around >3 μs (Fig. 4C), the lifetime of the proliferated skyrmion ~5 μs may not be simply following right after the cooling of transient temperature.

By more closely looking at the magnetic images, we can discuss the coalescence process occurring stochastically around the time range of 5±1 μs. As indicated in Fig. 4, A to C, we can define the time Regions I, II and III, according to the contour plot analysis (SOM, section 5). In Region I, three skyrmions (skA, skB1 and skB2) are well separated (Fig. 4F). In Region III, two large elliptical skyrmions (skA and skB) exist similarly as in the initial state (Fig. 4H). On the other hand, the contour plots in Region II show that these relevant skyrmions do exist but are connected or overlapped to each other (Fig. 4G). These observations can be explained by considering the event probability of respective skyrmion dynamics. As stated, in the present pump-probe measurements, 36-million events (20 kHz for 30 minutes acquisition time) are integrated into a single data at $t$. Therefore, the images presented here should represent the average of these 36-million events. Taking this into account, the plots in Region II should be interpreted as the weighted mean of those in Regions I and III just on the verge of Region II. Based on this consideration, we depict a schematic viewgraph showing the probability of the presence of proliferated skB1 and skB2 as a function of $t$ (Fig. 4I). We note that this $t$ dependence is also similar to the image in Fig. 4E.

From these results, we can understand the relaxation process as follows. Starting from the quasi-hexagonal skyrmions cluster at ~300 ns, skB1 moves back toward skB2 with slight elongation following the thermal relaxation down to RT (Fig. 4B). When the lower edges of skB1 and skB2 are aligned (~4 μs), the coalescence starts to occur stochastically in region II. Rapid decrease in the mean size of skB1 around 5±1 μs (Fig. 4E) can be interpreted as an exponential decay of the probability of the presence of proliferated skB1 and skB2 with μs time constant (Fig. 4I) which is attributed to weakened but still persistent topological protection around the local defects(*12*). Probability distribution averaged over 36 million events in Fig. 4I represents the repeatable nature of the skyrmion's whole life.

**Discussion**

We summarize our observations in Fig. 5. By irradiating the ns optical pulse, the flexible skyrmions found in the present setup behave as the following. (1) The ultrafast demagnetization occurs following the rapid heating (<1 ps). (2) Skyrmion proliferates in <1 ns. (3) Skyrmion contracts by shifting its position with a time delay of ~5 ns (damping), thus forming circular shape. (4) Quasi-hexagonal skyrmion cluster is formed in ~100 ns, *i.e.* self-organization, driven by the drifts of the reshaped skyrmions (orange circle in Fig. 5). (5) Hexagonal-skyrmion cluster maintains to ~300 ns and the relaxation process toward the disordered state follows. (6) Finally, the skyrmion coalescence occurs as a stochastic topological transition at 5±1 μs. We note that though these events are well resolved both in time and space, they should not be extrapolated to the general picture. Details of the dynamics, especially the lifetime of skyrmion, should strongly depend on materials, boundary conditions, temperature etc.

In the present $Co_9Zn_9Mn_2$ thin plate, we thus established the skyrmions' life cycle showing the successive disorder-order transformations. It is notable that the self-organization to the higher-symmetric skyrmion cluster is not instantaneously realized by the thermal excitation. Instead, the skyrmion-reshaping processes occur and then they start moving as particles, leading to the skyrmion cluster formation in a longer time scale. Such multiscale spatiotemporal structure akin to soft matters(*15,16*) is reasonably understood by the retarded responses related to the damping of internal spins, and directly reflects the hierarchy of internal microscopic spin magnetic moments and mesoscopic particle nature. The weakened but still persistent topological protection further adds the μs lifetime of the skyrmions. These results provide good initial insights into the kinetics of the flexible skyrmions around the imperfections, which inevitably exist in real materials and may limit the performance of the skyrmion-based spintronic devices(*13,14*). Further pump-probe LTEM experiments at lower fluences for investigating elastic dynamics of magnetic textures as well as the possible repeatable transformation between helical stripes and skyrmions will be important future works. The present results visualize the unique life cycle of the photothermally-induced metastable skyrmions, and also propose new concept for skyrmion-based spintronic applications based on the repeatable creation, annihilation and transportation of topological charge, such as high-frequency undulation of the emergent magnetic field in nanoscale spintronic circuit.

**Materials and Methods**

**Sample preparation**

Bulk $Co_9Zn_9Mn_2$ single crystals were synthesized using procedures described elsewhere(*25,26*). The thin plates for the LTEM observations were prepared from the bulk sample by using dual-beam focused-ion-beam (FIB) instrument (Hitachi, FB-5000). After the thinning process, we further irradiated the thin plate with Ga ion (acceleration voltage of 40 kV) at a grazing angle about 2 degrees to randomly introduce local crystal defects.

**Pump-probe Lorentz transmission electron microscopy**

Pump-probe LTEM measurements were performed at RIKEN CEMS (Thermo Fischer Scientific, Tecnai Femto). For the excitation of the sample, we used the laser pulse (1064 nm and 1 ns duration) delivered from a Q-switched laser operating at 20 kHz repetition (Bright Solutions, Wedge-HF). Its diameter at the sample position was set to 100 μm. For generating the electron packets, we used the laser pulse up-converted to 266 nm (10 ns duration) delivered from a Q-switched laser (Advanced Optowave, AWave-532). In this work, we obtained three datasets by pump-probe LTEM for the same sample. We used the dataset 1 (from −20 to 24 ns) for Fig. 2, E1 to E3, Fig.3, A to I, dataset 2 (from −430 to 3970 ns) for Fig. 2, C and D, E4 to E7, 3H, 3I, Fig. 4, A, B, and D, and the Figs. S2 and S3, and the dataset 3 (from −1280 to 15870 ns) for Figs. 2E8, Fig.4, A, B, E to H and the Fig. S4. We integrated the pulsed electrons for 30 minutes to obtain a LTEM image at a certain delay $t$.

**Acknowledgments:** We thank W. Koshibae, N. Nagaosa and M. Ishida for valuable discussion, K. Nakajima for experimental support. **Funding:** This work was supported by JSPS KAKENHI (grant numbers 18H01818, 19K22120 and 19H00660) and Japan Science and Technology Agency (JST) CREST program (Grant Number JPMJCR1874 and JPMJCR20T1). **Author contributions:** T.S. and K.I. conceived the project. T.S. and A.N. constructed and performed pump-probe LTEM measurements, and analyzed the data with input from X.Z.Y. X.Z.Y. analyzed LTEM images by using the transport-of-intensity equation. K.K. and Y.Taguchi synthesized $Co_9Zn_9Mn_2$ alloy. T.S. wrote the paper with inputs from all coauthors. **Competing interests:** Authors declare no competing interests. **Data and materials availability:** All data is available in the main text or the supplementary materials.

**Supporting Online Materials:**
Sections 1 to 5
Figs. S1 to S4

**Figures,**

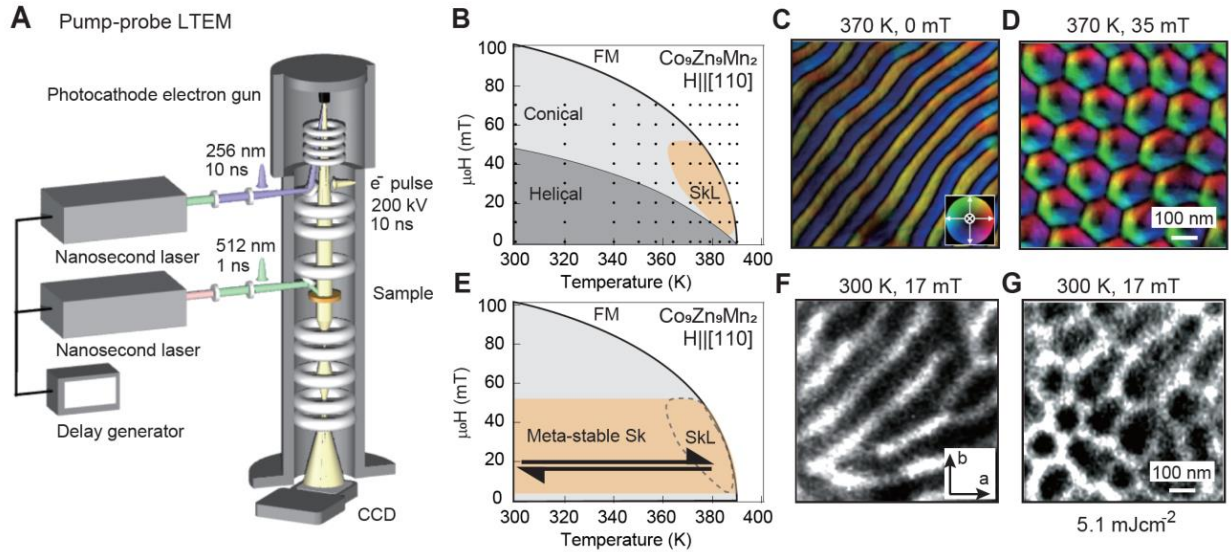

**Figure 1 Pump-probe Lorentz transmission electron microscopy on metastable skyrmions in $Co_9Zn_9Mn_2$.**
(**A**) Experimental geometry for the pump-probe LTEM. (**B**) Phase diagram of 100-nm thin-plate $Co_9Zn_9Mn_2$ obtained by LTEM measurements. Small black dots represent the data points. (**C** and **D**) Magnetization maps for helical stripes and equilibrium SkL obtained at 370 K in a magnetic field of 0 mT and 35 mT, respectively. (**E**) Phase diagram observed after the nanosecond laser irradiation with a fluence of 5.1 mJ/cm$^2$ (20 kHz for 5 seconds). Orange area indicates metastable skyrmion state, extended by the rapid field-cooling via the equilibrium skyrmion state (surrounded by black dotted curve) with the nanosecond laser irradiation. (**F** and **G**) LTEM images for thin plate with additional Ga-ion irradiation obtained at 300 K in a magnetic field of 17 mT before and after the laser irradiation, respectively. Crystal axes 110 and -110 are denoted as *a* and *b*, respectively.

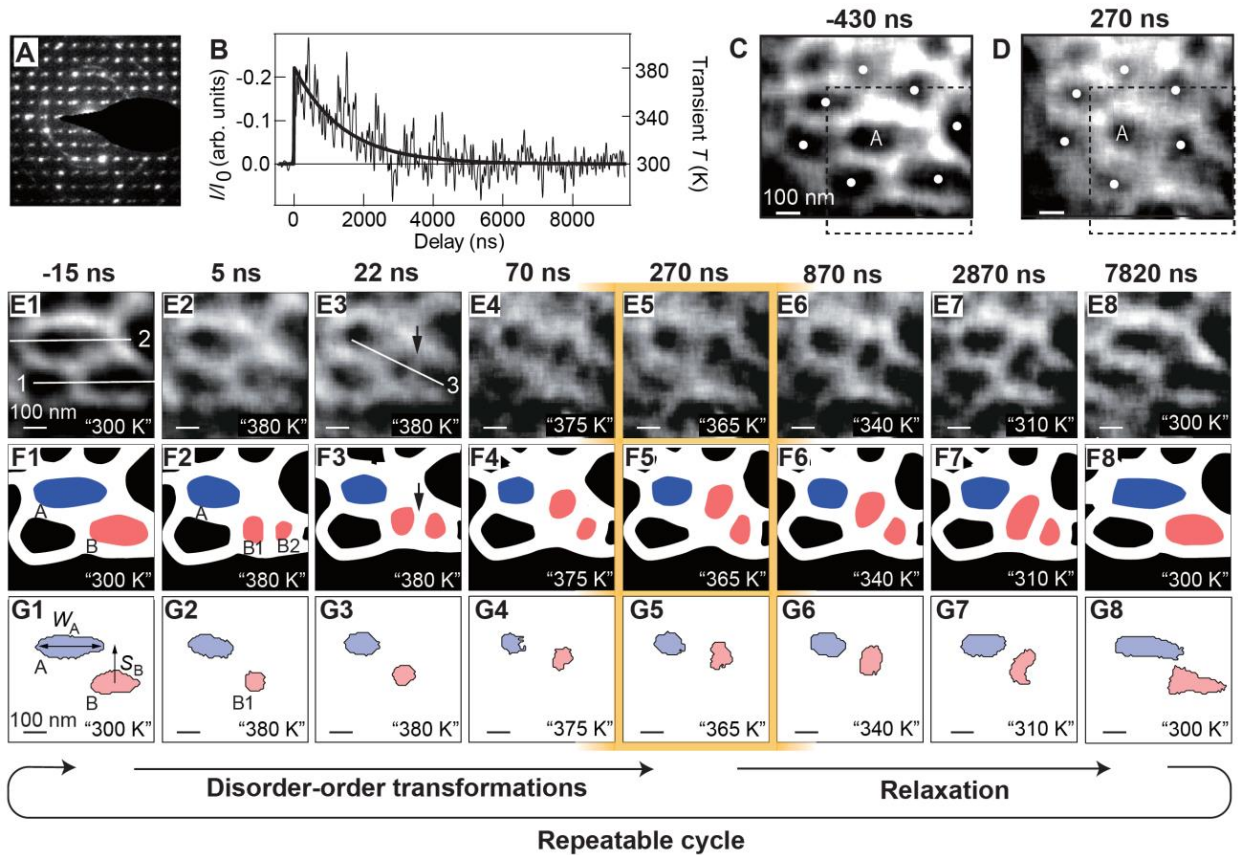

**Figure 2 One-cycle dynamics of metastable skyrmions**
(**A**) Electron diffraction pattern. (**B**) Time dependence of the Bragg spot intensity obtained by the pump-probe electron diffraction measurements. Rapid vibrations in the signal represent the bending motions of the thin film initiated by the laser irradiation. The thick black curve is the fitting result assuming an exponential relaxation with a time constant of 1.4 μs convoluted by the gaussian including the time resolution (10 ns). (**C** and **D**) LTEM images obtained at room temperature under 17 mT magnetic field before (–430 ns) and after (270 ns) the laser irradiation, respectively. Black dotted squares indicate the field of views of the LTEM images in (E). (**E1** to **E8**) LTEM images at –15 ns, 5 ns, 22 ns, 70 ns, 270 ns, 870 ns, 2870 ns and 7820 ns, respectively. (**F1** to **F8**) Schematics of the LTEM images in (E). (**G1** to **G8**) Contour plots representing the morphology of the skyrmions in (E). Transient temperatures obtained from the thick black curve in (B) are indicated at the bottom of the panels. Orange region in (E, F and G) represents the time scale required to form the quasi-hexagonal skyrmion cluster.

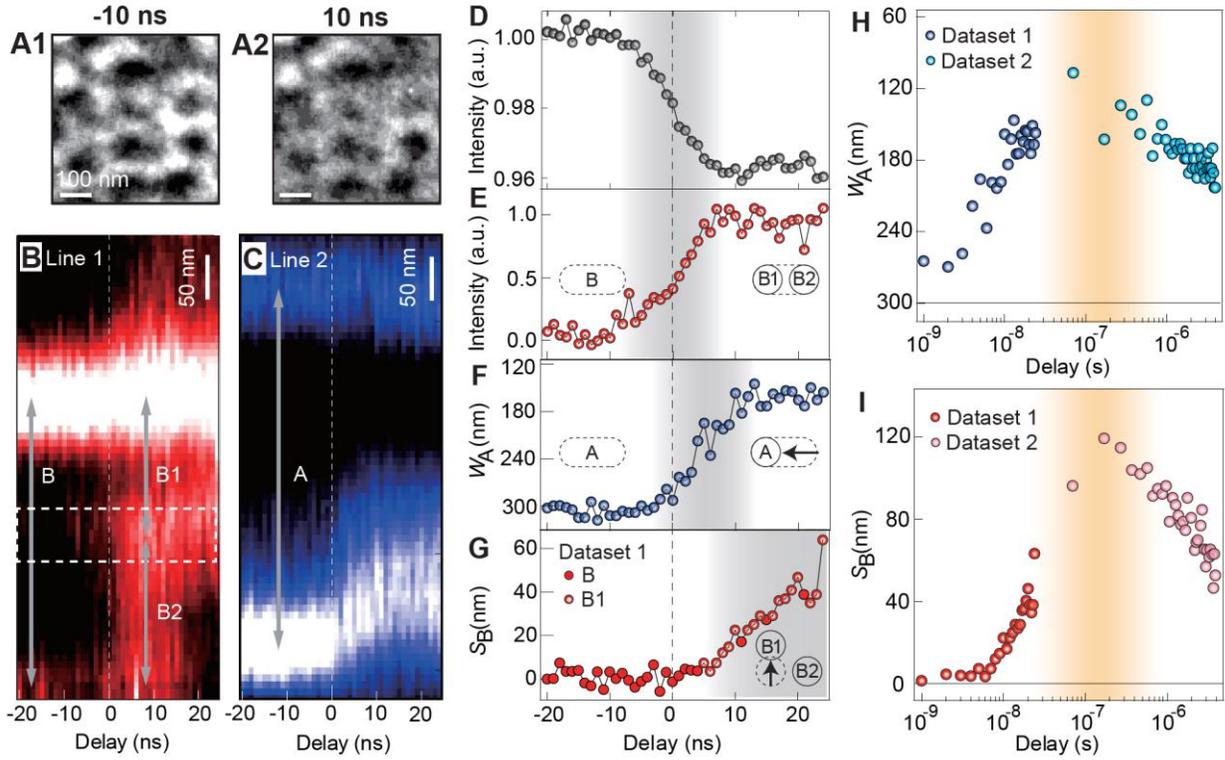

**Figure 3 Skyrmion proliferation, contraction and self-organization.** (**A1** and **A2**) LTEM images at −10 ns and 10 ns, respectively. Decrease in the magnetic contrast indicates a demagnetization due to the thermal excitation from 300 to 380 K. (**B**) Time dependence of the intensity profile along line 1 in Fig. 2E1. (**C**) The same as (B) but along line 2 in Fig. 2E1. (**D**) Time dependence of the integrated intensity of the LTEM images in (A). (**E**) Time dependence of the LTEM intensity in (B) (white dotted rectangle) showing the emergence of the magnetic wall between skB1 and skB2. (**F** and **G**) Time dependences of the width along $a$ of skA ($W_A$) and the shift along $b$ for skB (filled circles) and skB1 (open circles) ($S_B$), respectively, estimated from the contour plots in Fig. 2G. (**H**) The same as (F) but on a longer time scale. Blue and light blue markers are obtained from the dataset 1 (from −20 to 24 ns) and 2 (from −430 to 3970 ns), respectively. (**I**) The same as (G) but on a longer time scale. Red and pink markers are obtained from the dataset 1 and 2, respectively. The orange areas in (H and I) represent the time scale to form the quasi-hexagonal skyrmion cluster.

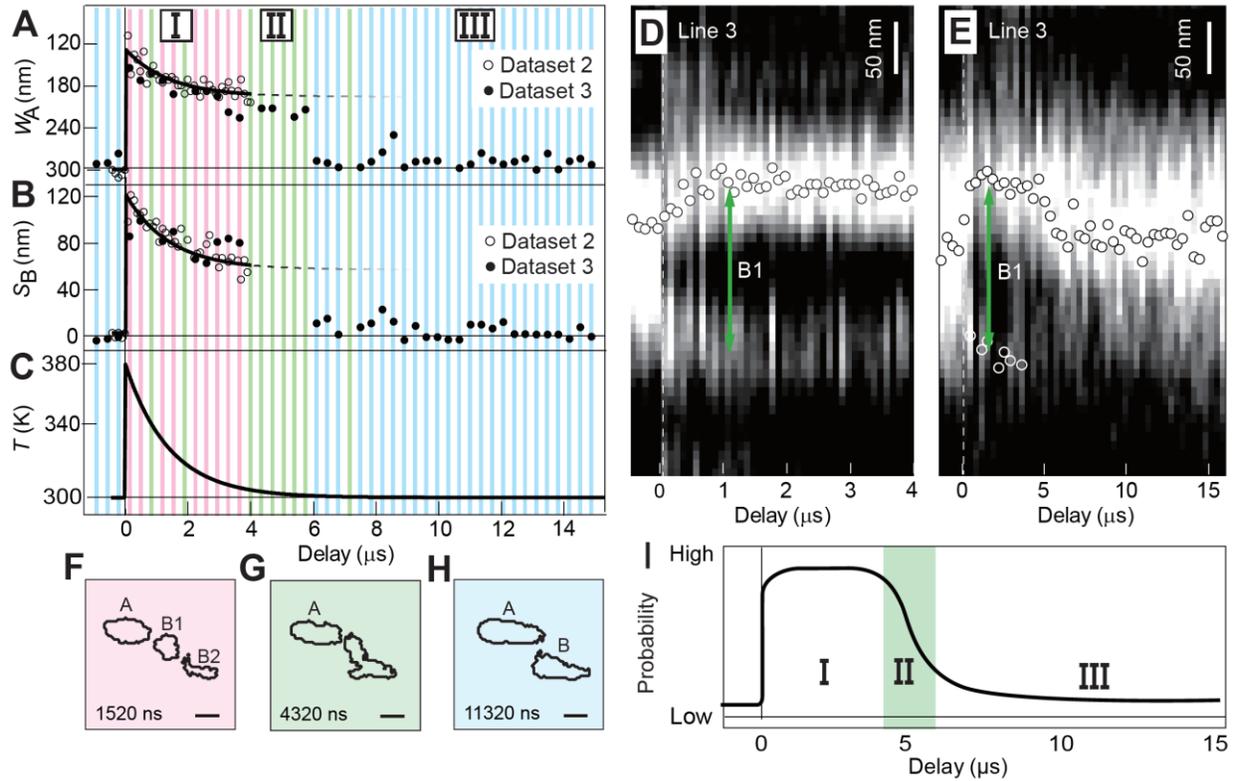

**Figure 4 Relaxation process and skyrmion coalescence.** (**A** and **B**) Time evolutions of $W_A$ and $S_B$ on a linear time scale. Open and filled markers are obtained from the dataset 2 (from –430 to 3970 ns) and 3 (from –1280 to 15870 ns), respectively. The black curves in (A and B) represent the fitting functions assuming an exponential decay with a time constant of 1.4 µs and a constant background (SOM, section 4). (**C**) Transient temperature deduced from the electron diffraction data in Fig. 2B. (**D**) Time dependence of the intensity profile along line 3 in Fig. 2E3 obtained from the dataset 2. (**E**) The same as (D) but on a longer time scale obtained from the dataset 3. (**F** to **H**) Typical contour plots for the LTEM images in the region I, II and III, respectively. Color bars in (A to C) represent the three characteristic states shown in (F to H) obtained from the dataset 3. (**I**) Schematics for the probability of the presence of proliferated skB1 and skB2 as a function of $t$.

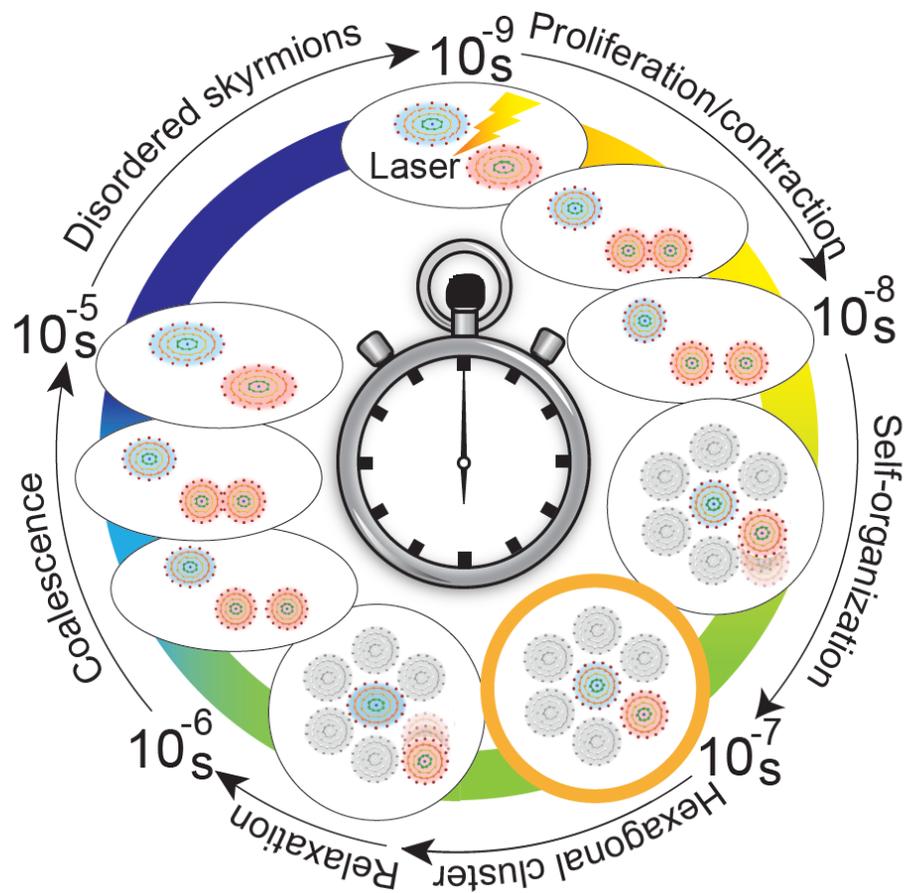

**Figure 5 Repeatable life cycle of the magnetic skyrmions in the present setup.** Blue, red and gray wheels represent the skA, skB/skB1/skB2 and skyrmions surrounding skA and skB, respectively. Orange circle represents the time scale required to form the quasi-hexagonal skyrmion cluster.

# Supporting online materials for
# Nano-to-micro spatiotemporal imaging of
# magnetic skyrmion's life cycle

T. Shimojima[1], A. Nakamura[1], X. Z. Yu[1], K. Karube[1], Y. Taguchi[1], Y. Tokura[1,2,3], K. Ishizaka[1,2]

[1]RIKEN Center for Emergent Matter Science (CEMS), Wako 351-0198, Japan
[2]Quantum-Phase Electronics Center (QPEC) and Department of Applied Physics, The University of Tokyo, Tokyo 113-8656, Japan
[3] Tokyo College, The University of Tokyo, Tokyo 113-8656, Japan

**This PDF file includes:**
Sections 1 to 5
Figs. S1 to S4

1. **Estimation of the temperature jump by ns-laser irradiation.**

We investigated laser fluence ($F$) and magnetic field ($B$) dependences of the LTEM images in order to estimate the temperature jump of the thin film after the nanosecond laser irradiation. The measurement procedure is as follows. First, we set $B = B_0$ and irradiated ns laser with $F = F_0$. Then, we obtained a LTEM image by thermionic electron beam with an exposure time of 1 second under the nanosecond laser irradiation. The repetition rate of the laser was set to 20 kHz which induces 20-thousands temperature cycles in 1 second. After resetting the magnetic field ($B_0 \rightarrow 2\,\text{T} \rightarrow B_0$), we repeat the data acquisition with increased $F$ ($F_0 \rightarrow F_1$). We then obtain a series of $F$-dependent LTEM images at $B_0$. After resetting and increasing the magnetic field ($B_0 \rightarrow 2\,\text{T} \rightarrow B_1$), we repeat the same $F$-dependent measurements at $B_1$. Accordingly, we obtained 663 LTEM images in a wide $B$-$F$ region ($0 \leq B \leq 92$ mT and $0 \leq F \leq 6.3$ mJcm$^{-2}$) as indicated by the data points in Fig. S1A (white dots).

In Fig. S1A, we show the $B$-$F$ phase diagram showing the number of the skyrmions ($N_{sk}$) in a selected field of view of the LTEM images. Note that the LTEM images in Fig. S1, B to E, tend to show a stripe magnetic structure parallel to $b$ axis especially in high $B$, due to the sample tilting angle of ~5 degrees with respect to the external magnetic field. Here, we found that the helical (H) structures transform into the disordered skyrmions above a critical fluence $F_{c1}$ ($4.3 < F_{c1} < 4.7$ mJcm$^{-2}$ at 17 mT, $3.9 < F_{c1} < 4.3$ mJcm$^{-2}$ at 32 mT and 41 mT). By further increasing $F$, $N_{sk}$ suddenly decreased above $F_{c2}$ ($5.5 < F_{c2} < 5.9$ mJcm$^{-2}$ at 32 mT, $4.7 < F_{c2} < 5.1$ mJcm$^{-2}$ at 41 mT). We found that both $F_{c1}$ and $F_{c2}$ are clearly dependent of $B$, and the metastable skyrmions never appear above 53mT.

Emergence of the skyrmions at $F_{c1} < F < F_{c2}$ indicates the formation of metastable skyrmions by the field cooling through the equilibrium skyrmion state. On the other hand, absence of the skyrmion at $F > F_{c2}$ suggests that the sample temperature exceeds the transition temperature between equilibrium skyrmion state and ferromagnetic (FM) state. Once the sample is heated up to the FM state, the magnetic contrast should completely disappear. In this case, it might be difficult to create the metastable skyrmions at the same arrangement for each temperature cycle. As a result, the metastable skyrmions cannot appear in the LTEM image which is averaged by 20-thousands temperature cycles.

Based on these considerations, we can overlay the skyrmion-FM phase boundary of the thin plate (white broken curve in Fig. S1A) on the $F_{c2}$ trajectory in the obtained $B$-$F$ phase diagram. Then, we can estimate the temperature jump as a function of $F$, as indicated by the correspondence between the top and bottom axes of Fig. S1A. Noting that the bright area in the $B$-$F$ phase diagram well corresponds to equilibrium skyrmion state of the thin film (orange area in Fig. 1B). The ns-laser irradiation of $F = 5.1$ mJcm$^{-2}$ in the main text is thus expected to increase the sample temperature up to ~380 K.

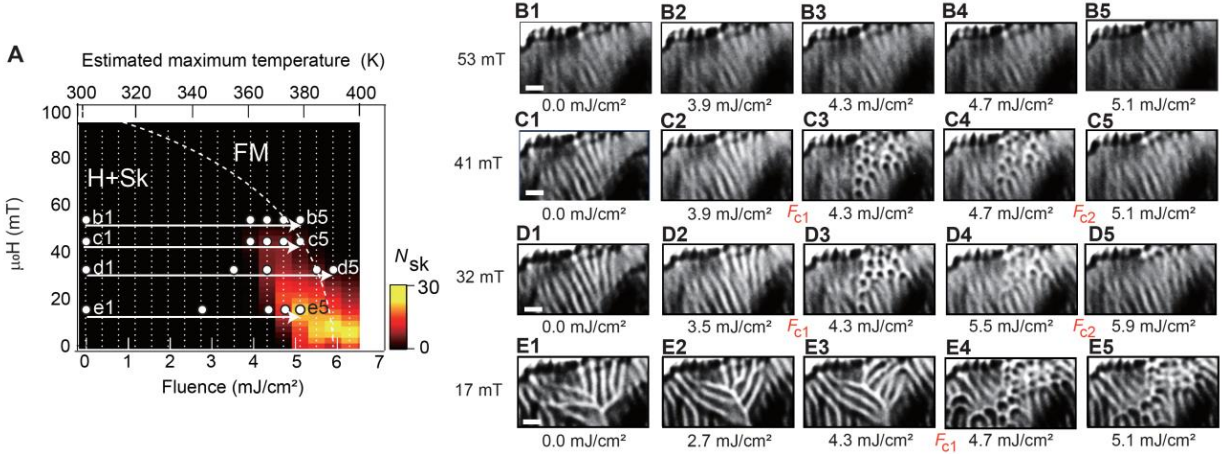

**Fig. S1| Estimation of the temperature jump with ns-laser irradiation.** (**A**) *B-F* phase diagram showing the number of skyrmions ($N_{sk}$) in the LTEM images in **b-e**. White small dots represent the data points. Filled white markers represent the selected data points corresponding to (B to E). White broken curve indicates the phase boundary between skyrmion/helical and FM states for the thin-plate sample taken from Fig. 1B. (**B to E**) *F*-dependences of the LTEM images under 53 mT, 41 mT, 32 mT and 17 mT magnetic field, respectively.

2. **Reduced lifetime of the deformed and disordered skyrmions.**
We note that the repeatable skyrmion dynamics in Fig. 5 was observed only in a certain area of the thin film where the Ga ion was additionally irradiated after the thinning process by the FIB method. This observation suggests the presence of the spatial variation in the skyrmion lifetime. It also indicates that there should be some local effect of inhomogeneity that affects the "topological protection" of skyrmions (i.e. the continuity of the spin texture) and reduces the lifetime to < 50 μs (shorter than the period of the pump-probe cycle). These observations are understood by the influences from the crystal defects intentionally introduced. For the thin film without additional Ga ion irradiation, we could not observe deformed and disordered skyrmions, and their repeatable dynamics.

3. **Contour plot analysis of the LTEM images.**
We obtained the contour plots from the LTEM images by employing adaptive thresholding. We binarize the LTEM image using this method which calculates the thresholds in selected regions ($n \times n$ pixels) surrounding each pixel. Each threshold value is obtained by the weighted mean of the local area minus an offset value, which defines the location of the boundary between high and low intensity regions in the LTEM images. Changes in the *n* and offset values give rise to slight difference in the skyrmion shape. We chose three different sets of these values for the datasets 1, 2 and 3 for clearly tracking the motion and transformation of skyrmions in each dataset. We carefully checked that the contour plots are intuitively consistent with the raw LTEM images and schematics shown in Fig. 2, E and F.

    Here we describe the details of the data analysis for the estimation of size and position of skyrmions. Figure S2 shows the definitions of the width along *a* of skA ($W_A$) and the shift along *b* of skB ($S_B$) in the contour plots at −430 ns and 870 ns. First, we put the boxes which cover the edges of skA and skB. Second, the $W_A$ and $S_B$ were estimated as indicated by the double-headed arrows and the arrow in Fig. S2, respectively. For these analyses, the error bars are smaller than the size of the markers in Figs. 3 and 4.

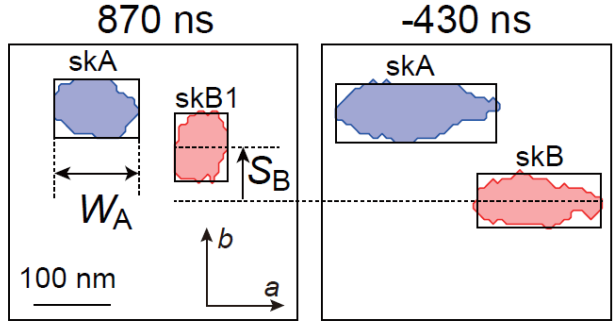

**Fig. S2 | Definitions of the width along *a* for skA ($W_A$) and the shift along *b* for skB ($S_B$).**

### 4. Fitting analysis for transient $W_A$ and $S_B$ values.

We reproduced the relaxation processes of $W_A$ and $S_B$ in the time region of <4 μs for the dataset 2 by assuming the exponential decay of transient temperature with time constant of 1.4 μs, as indicated by the red curves in Fig. S3, A and B, respectively. The constant background (green) was also included in order to reproduce the flat feature around 4 μs in the data, indicating the influence from the topological protection in skB1 and skB2.

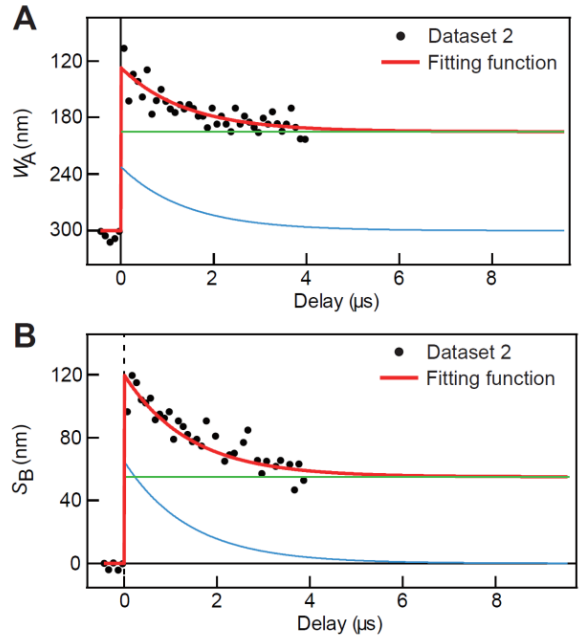

**Fig. S3 | Fitting analysis for transient $W_A$ and $S_B$ values**. (**A**, **B**) Fitting analysis on $W_A$ and $S_B$ for the dataset 2, respectively. Red curves consist of an exponential component with a time constant of 1.4 μs (blue curves) and a constant background (green lines).

### 5. Contour plot analysis for the dataset 3

We show the contour plots with boxes for the dataset 3 in Fig. S4, which were used to define the time regions I, II and III in Fig. 4, A to C. The colors correspond to the three types of the contour plots as indicated in Fig. 4, F to H (red for region I, green for region II and blue for region III).

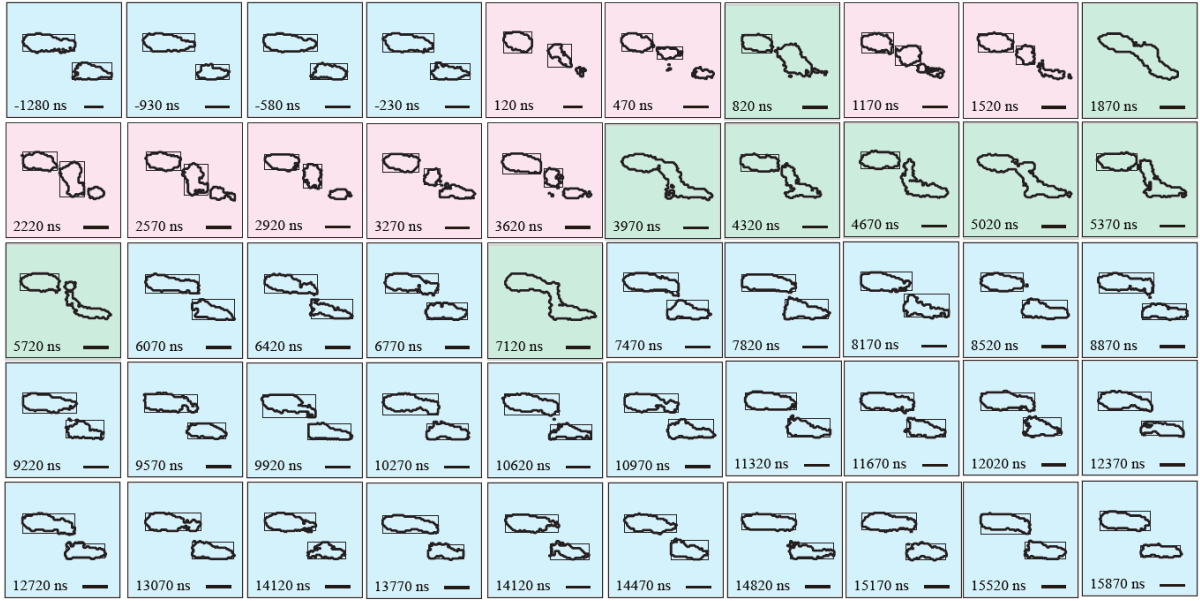

**Fig. S4 | Contour plots for the dataset 3**.